\documentclass[11pt,a4paper]{article}

\usepackage{graphicx}
\usepackage{bm}
\usepackage{amsmath} 
\usepackage{ragged2e} 
\usepackage{placeins} 

\usepackage{caption}
\usepackage{subcaption}
\usepackage[labelfont=bf]{caption}
\usepackage{hyperref}
\hypersetup{
  colorlinks=true,
  linkcolor=black,
  citecolor=blue,
  urlcolor=blue
}

\usepackage[
  left=2.5cm,
  right=2.5cm,
  top=2.5cm,
  bottom=2.5cm
]{geometry}

\usepackage[backend=biber,style=numeric,sorting=none]{biblatex}
\def\papertitle{On the interpretation of Hahn echo measurements in electron spin resonance scanning tunneling microscopy}

\newcommand{\AuthorList}{\normalsize
    Paul Greule$^1$, Wantong Huang$^1$, Máté Stark$^1$, Kwan Ho Au-Yeung$^{1,2}$, Christoph Wolf$^{3,4}$,\\ \normalsize Soo-hyon Phark$^{3,4}$, Andreas J. Heinrich$^{3,4}$ and Philip Willke$^{1,2*}$
}

\newcommand{\Affiliations}{
\small
\noindent $^{1}$Physikalisches Institut (PHI), Karlsruhe Institute of Technology (KIT), Karlsruhe, Germany.\\
$^{2}$Center for Integrated Quantum Science and Technology (IQST), Karlsruhe Institute of Technology, Karlsruhe, Germany.\\
$^3$Center for Quantum Nanoscience, Institute for Basic Science (IBS), Seoul 03760,\\ Republic of Korea.\\
$^4$Ewha Womans University, Seoul 03760, Republic of Korea.\\
\normalsize
}

\newcommand{\CorrespondingAuthor}{%
\small $^{*}$corresponding author: philip.willke@kit.edu \normalsize
}

\title{\bfseries \boldmath \papertitle}
\author{\AuthorList}
\begin{document}
\date{}
\maketitle
\Affiliations \\
\CorrespondingAuthor

\vspace{1cm}
\begin{abstract}
\justifying
\noindent \textbf{Electron spin resonance scanning tunneling microscopy (ESR-STM) has become a powerful tool for probing spin dynamics and coherence of individual atoms and molecules on surfaces. In this work, we perform Rabi oscillation and Hahn echo pulse protocols on individual iron phthalocyanine (FePc) molecules on MgO/Ag(001) using ESR-STM. While Hahn echo protocols are widely used to extract spin coherence times, we show that in ESR-STM they are particularly susceptible to misinterpretation due to tunneling electrons generated by the applied radio-frequency (RF) voltage. The RF voltage not only drives the spin, but simultaneously probes and relaxes it, which consequently leads to an exponential decay that reflects spin relaxation ($T_1$) rather than intrinsic phase coherence ($T_2$). We moreover show that varying both delay times in the refocusing pulse sequence is a reliable way to ensure a coherent nature of the echo signal. The extracted decay for the latter protocol suggests that $T_2^{\mathrm{Hahn}}$ is ~$\approx 30~\mathrm{ns}$ and is thus closer to the decoherence time $T_2^{\mathrm{Rabi}}$ observed in Rabi oscillation measurements. This is significantly shorter than values reported in previous echo measurements. Our findings underscore the need for caution in interpreting $T_2$ times from Hahn echo and Carr-Purcell protocols in ESR-STM and provide practical criteria for distinguishing true spin echoes from tunneling-induced relaxometry signals.}
\end{abstract}

\FloatBarrier
\newpage
\section{Introduction}
The combination of electron spin resonance (ESR) and scanning tunneling microscopy (STM) has developed into a powerful technique with high spatial, time and energy resolution\cite{baumann_electron_2015,chen_harnessing_2023}. By coupling a radio-frequency (RF) voltage to the STM junction, ESR-STM enables coherent control of single spins, which has been demonstrated for both adatoms, such as Ti\cite{yang_coherent_2019,wang_universal_2023,wang_atomic-scale_2023}, and molecules\cite{willke_coherent_2021,kovarik_spin_2024,huang_quantum_2025}, including iron phthalocyanine (FePc), on surfaces. These developments establish ESR-STM as a versatile platform for exploring quantum coherence at the atomic scale.

Pulsed ESR-STM experiments have adopted a range of established spin-control protocols originally developed in magnetic resonance\cite{slichter_principles_2013} and quantum information science\cite{degen_quantum_2017, childress_coherent_2006}. These include Rabi oscillations and Ramsey interferometry\cite{yang_coherent_2019}, Hahn echo protocols\cite{yang_coherent_2019,willke_coherent_2021,wang_atomic-scale_2023}, and extended refocusing sequences such as Carr–Purcell protocols\cite{wang_universal_2023}. In conventional magnetic resonance, Hahn echo sequences are used to refocus inhomogeneous dephasing and to access intrinsic phase coherence times by suppressing reversible contributions to decoherence.

In ESR-STM, echo-based protocols have similarly been employed with the goal to improve and understand spin coherence. Hahn echo decoherence times $T_{\mathrm{2}}^{{\mathrm{Hahn}}}$ on the order of several tens to a few hundred nanoseconds for both atomic and molecular spins have been reported\cite{yang_coherent_2019,willke_coherent_2021,wang_universal_2023}.

Coherent spin control is achieved by applying RF voltage pulses $V_{\mathrm{RF}}(t)$ on a nanosecond timescale to the STM tip in addition to a static bias $V_{\mathrm{DC}}$, allowing spin rotations while maintaining a finite tunneling current. 
For  readout and control via the STM tip, understanding the microscopic mechanism has motivated several experimental and theoretical works. For instance, how the RF voltage can drive the spin state has gained significant attention: Early models emphasized electric-field-driven or piezoelectric mechanisms\cite{baumann_electron_2015,lado_exchange_2017}, while more recent work has identified electron-transport-mediated excitation as a dominant contribution in many systems\cite{reina-galvez_contrasting_2025,reina-galvez_efficient_2025,kovarik_spin_2024}.

For the detection of the spin state in ESR-STM the change in tunneling current $\Delta I$ through the junction is used. The latter is influenced by both the static DC bias and the applied RF voltage. RF-induced tunneling currents have been shown to probe the spin state via a homodyne-like detection mechanism\cite{yang_coherent_2019}, and for molecular spins the detection channel has been discussed in terms of spin-transfer torque and field-like torque effects\cite{kovarik_spin_2024}.
As a consequence, the RF voltage excitation in ESR-STM simultaneously acts as a coherent drive as well as a probe of the spin system. In addition, the tunneling currents, induced by $V_{\mathrm{RF}}$ and $V_{\mathrm{DC}}$ have also been shown to introduce spin relaxation and decoherence\cite{yang_coherent_2019,paul_control_2017,willke_coherent_2021}. Besides a direct coherent control of a spin system underneath the STM tip, also coherent control of remote spins has been realized\cite{wang_atomic-scale_2023}. Here, the readout is not performed via the change in tunneling current directly, but via a dispersive readout using a second atom as a readout spin. 

For a direct readout scheme, the intertwined role of driving, detection, and relaxation requires careful interpretation of pulsed ESR-STM experiments.
In this work, we demonstrate how the role of the $V_{\mathrm{RF}}$ voltage affects the results of Hahn echo experiments. Using FePc molecules on magnesium oxide (MgO) on Ag(001) as a model system, and drawing connections to established results, we investigate Hahn echo protocols in greater detail. We find that they can yield exponential decay signals that resemble coherent echo decays but may instead reflect incoherent spin relaxation processes. As a result, decay times extracted from such measurements can significantly overestimate the intrinsic phase coherence time. We further demonstrate that systematically varying the first and second delay time in the Hahn echo sequence provides a stringent test of the genuine echo.

\FloatBarrier
\newpage
\section{Experimental Setup and Rabi Characterization}
All measurements were performed in a home-built low-temperature scanning tunneling microscope setup equipped with a commercial STM head (UNISOKU USM-1600) and operated at a base temperature of approximately $50~\mathrm{mK}$. An external magnetic field $B$ was applied perpendicular to the sample surface to define the spin quantization axis.

\begin{figure*}[t]
	\centering
	\includegraphics[width=120mm]{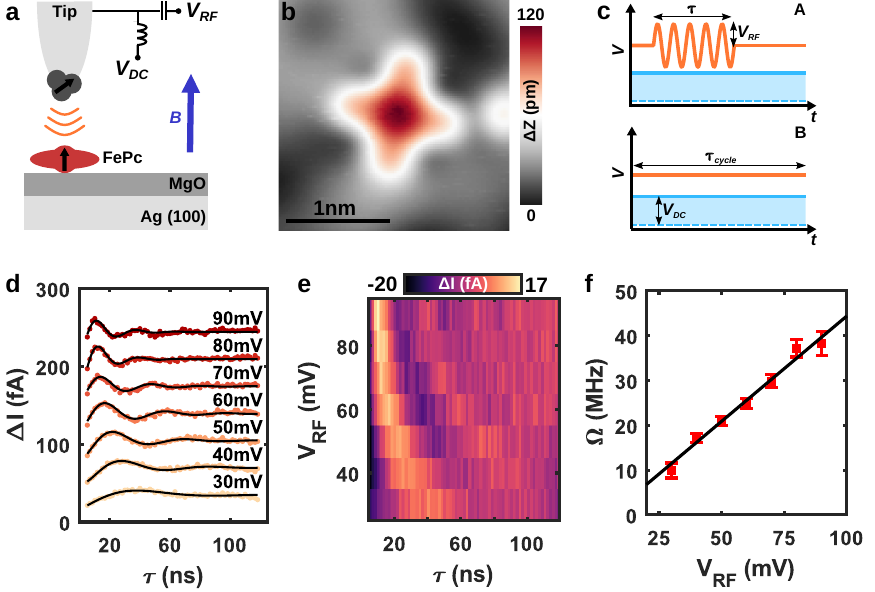}
  \caption{
  \justifying
  \textbf{Rabi oscillation measurements of FePc on MgO/Ag(001).} 
  \textbf{(a)} Schematic of the ESR-STM setup. An RF voltage $V_{\mathrm{RF}}$ is applied to the STM junction in addition to a constant bias $V_{\mathrm{DC}}$, while an out-of-plane magnetic field $B$ defines the quantization axis. \textbf{(b)} STM topography of an FePc molecule adsorbed on a bilayer MgO/Ag(001) surface ($I_{\mathrm{set}}=50~\mathrm{pA}$, $V_{\mathrm{DC}}=100~\mathrm{mV}$). \textbf{(c)} Pulse sequence used in the A and B cycle of the lock-in scheme for Rabi oscillation measurements. The spin is coherently driven by $V_{\mathrm{RF}}$ for a pulse duration $\tau$, while a background voltage $V_{\mathrm{DC}}$ is constantly applied. \textbf{(d)} Experimental Rabi oscillation measurements: Change in current $\Delta I$ as a function of $\tau$ for different $V_{\mathrm{RF}}$ amplitudes (30--90 mV). The traces show oscillatory behavior
  characteristic of coherent spin rotations. Solid lines represent fits to $\Delta I(t)=I_0 \sin(\Omega t+\phi)\exp(-t/T_2^{\mathrm{Rabi}})$. Traces are vertically offset for clarity. A linear background due to current rectification ($k_{\mathrm{lin}}$) has been subtracted ($I=5~\mathrm{pA}$, $V_{\mathrm{set}}=
  40~\mathrm{mV}$, $B=536~\mathrm{mT}$, $f_0=14.772~\mathrm{GHz}$, $\tau_{\mathrm{cycle}}=800~\mathrm{ns}$). \textbf{(e)} Colormap of the data shown in (d). \textbf{(f)} Extracted Rabi frequency $\Omega$ as a function of $V_{\mathrm{RF}}$ from the data in (d) with  a linear dependence (black line; slope: $0.47~\mathrm{MHz/mV}$).}
  \label{fig:fig1}
\end{figure*}

Figure~\ref{fig:fig1}a shows a schematic of the ESR-STM setup. In addition to a static bias voltage $V_{\mathrm{DC}}$, a radio-frequency voltage $V_{\mathrm{RF}}$ is coupled to the STM junction via the tip, enabling coherent spin manipulation while maintaining a finite tunneling current $I$. Spin readout is performed electrically through changes in the tunneling current induced by the spin-dependent conductance of the junction. The spin system investigated in this work consists of individual FePc molecules adsorbed on a bilayer of MgO grown on Ag(001). An STM topography of a single FePc molecule is shown in Fig.~\ref{fig:fig1}b. FePc on
MgO/Ag(001) has previously been established as an isotropic spin-$1/2$ system via ESR-STM\cite{zhang_electron_2022,willke_coherent_2021}. Coherent spin control and pulsed ESR-STM protocols have been demonstrated on this system, making it a suitable platform to investigate pulse sequences.
Figure~\ref{fig:fig1}c shows the pulse sequence used for Rabi oscillation
measurements, adapted from Ref.~\cite{yang_coherent_2019}. Spin readout is
performed via a lock-in detection scheme that compares the time-averaged
tunneling current between two pulse cycles, where the RF voltage is only present in the A cycle, while  the DC voltage is kept on continuously. This yields a measured current
difference $\Delta I = \overline{I}_A - \overline{I}_B$. The detected signal can be described as [Eq. (S10) in Ref.\cite{yang_coherent_2019}]

\begin{align}
\Delta I \propto {}{}&
k_{\mathrm{DC}}\left[-\sin \Omega t + \Omega T_1 \left(1-\cos \Omega t\right)\right]
+  k_{\mathrm{RF}} \left(1-\cos \Omega t\right) + k_{\mathrm{lin}}\, t
\label{eq:deltaI}
\end{align}
\\
\noindent where $\Omega$ is the Rabi frequency, i.e. the speed with which the spin is rotated between ground and excited state. $T_1$ is the spin relaxation time and the coefficients $k_{\mathrm{DC}}$, $k_{\mathrm{RF}}$, and $k_{\mathrm{lin}}$ account for contributions from DC-bias-driven tunneling ($k_{\mathrm{DC}}$), RF-induced tunneling currents ($k_{\mathrm{RF}}$) and rectification effects ($k_{\mathrm{lin}}$), respectively, and are given by
\begin{align}
k_{\mathrm{DC}}
&= \frac{G_{\mathrm{junc}}\,V_{\mathrm{DC}}}{2\,\Omega\,\tau_{\mathrm{cycle}}}\,
   \langle S_t^{z}\rangle, \\
k_{\mathrm{RF}}
&= \frac{G_{\mathrm{junc}}\,V_{\mathrm{RF}}}{4\,\Omega\,\tau_{\mathrm{cycle}}}\,
   \sin\phi_{\mathrm{RF}}\,\langle S_t^{x}\rangle, \\
k_{\mathrm{lin}}
&= \frac{G_{\mathrm{junc}}\,V_{\mathrm{DC}}}{2\,\tau_{\mathrm{cycle}}}\,
   \langle S_t^{z}\rangle + k_{\mathrm{rec}}
\end{align}
\\
\noindent Here, \(G_{\mathrm{junc}}\) denotes the spin-averaged conductance of the tunnel junction and \(\tau_{\mathrm{cycle}}\) the pulse repetition period. The phase \(\phi_{\mathrm{RF}}\) accounts for the phase difference between the applied RF voltage and the precession of the spin in the laboratory frame. $k_{\mathrm{rec}} $ accounts moreover for RF rectified background currents due to the non-linearity in the I-V curve. The quantities \(\langle S_t^{x} \rangle\) and \(\langle S_t^{z} \rangle\) denote the time-averaged transverse and longitudinal components of the magnetic tip spin. A key aspect of Eq.~\eqref{eq:deltaI} is the contribution proportional to $k_{\mathrm{RF}}$, which allows the spin state to be probed by tunneling electrons generated solely by the RF voltage, even in the absence of a DC bias.
This homodyne-like detection mechanism depends on the orientation of the tip magnetization relative to the surface spin and is maximized for an in-plane tip magnetization \(\langle S_t^{x} \rangle\).~\cite{yang_coherent_2019}

Figures~\ref{fig:fig1}d--f present Rabi oscillation measurements on FePc, revealing a linear dependence of the Rabi frequency $\Omega$ on $V_{\mathrm{RF}}$ and a coherence time $T_2^{\mathrm{Rabi}} \approx 30~\mathrm{ns}$, both in good agreement with earlier work~\cite{willke_coherent_2021}. 

\FloatBarrier
\section{Hahn Echo and Carr--Purcell Sequences}
\begin{figure*}[t]
	\centering
	\includegraphics[width=144mm]{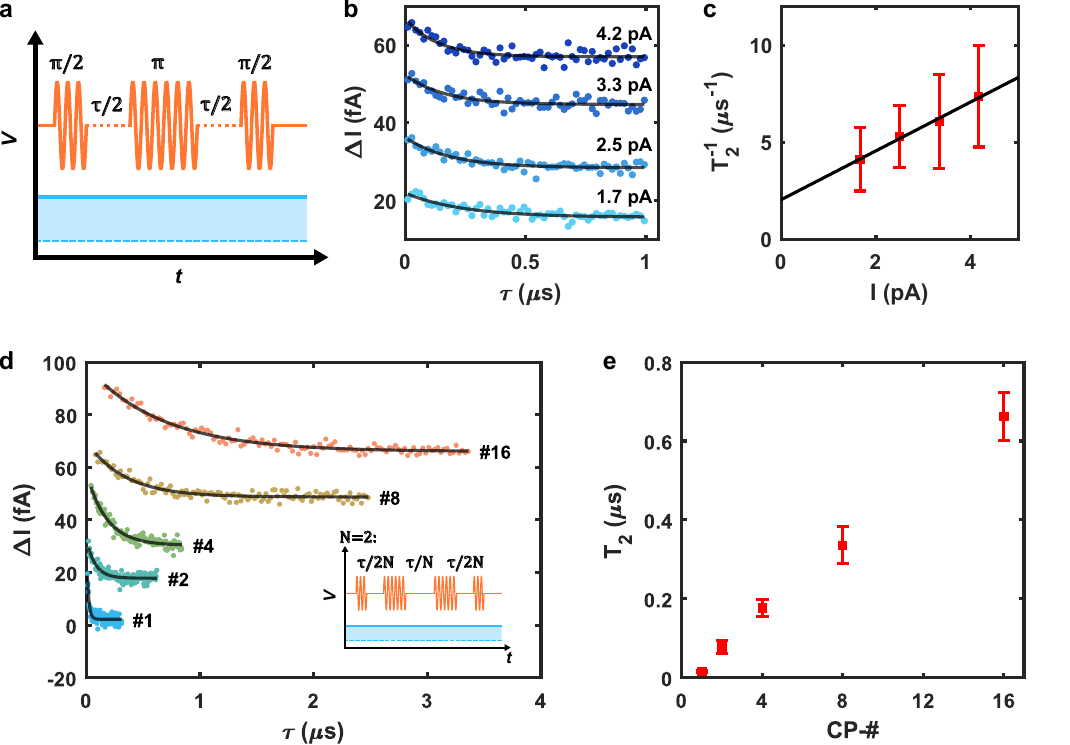}
  \caption{
  \justifying
  \textbf{Hahn echo and Carr--Purcell (CP) measurements.} \textbf{(a)} Pulse sequence used for Hahn echo, consisting of two $\pi/2$ pulses separated by a $\pi$ pulse and a background DC voltage (blue). \textbf{(b)} Change in tunneling current $\Delta I$ for a Hahn echo sequence as a function of pulse separation $\tau$ for different tunneling currents ($V_{\mathrm{DC}}=60~\mathrm{mV}$, $V_{\mathrm{RF}}=60~\mathrm{mV}$, $B=523~\mathrm{mT}$, $f_0=13.813~\mathrm{GHz}$, $\tau_\pi=15.9~\mathrm{ns}$, $\tau_{\mathrm{cycle}}=2500~\mathrm{ns}$). Solid lines represent exponential fits of the form $\Delta I(\tau)=I_0 \exp\!\left(-\tau/T_2^{\mathrm{Echo}} \right)$. Datasets are offset for clarity. \textbf{(c)} Extracted echo decay rates $T_2^{-1}$ as a function of tunneling current $I$, showing a linear dependence. \textbf{(d)} CP-measurements with increasing numbers of refocusing $\pi$ pulses (CP-1 to CP-16). The inset illustrates the CP-$N$ sequence for $N=2$ composed of an initial $\pi/2$ pulse, $N$ evenly spaced $\pi$ pulses with delay time $\tau/N$, and a final $\pi/2$ pulse ($I=7~\mathrm{pA}$, $V_{\mathrm{DC}}=50~\mathrm{mV}$, $V_{\mathrm{RF}}=60~\mathrm{mV}$, $B=536~\mathrm{mT}$, $f_0=13.991~\mathrm{GHz}$, $\tau_\pi=8.6~\mathrm{ns}$. $\tau_{\mathrm{cycle}}=0.8 \text{--} 7.6~\mathrm{\mu s}$). \textbf{(e)} Extracted coherence time $T_2$ as a function of the number of $\pi$ pulses in the CP sequence, revealing an apparent linear increase in $T_2$ with pulse number.}
	\label{fig:fig2}
\end{figure*}

From these measurements, we determine the pulse durations required for $\pi/2$ and $\pi$ rotations used in the Hahn echo scheme (Fig.~\ref{fig:fig2}a).

Hahn echo sequences performed on FePc are shown in Fig.~\ref{fig:fig2}b for different tunneling currents $I$. The Hahn echo sequence consists of an initial $\pi/2$ pulse that rotates the spin into a superposition of states $\lvert 0 \rangle$ and $\lvert 1 \rangle$, followed by a refocusing $\pi$ pulse and a second $\pi/2$ pulse that converts the refocused spin state into the measurement axes. In conventional magnetic resonance, this protocol suppresses inhomogeneous phase noise from static or slowly varying field fluctuations and enables access to the intrinsic phase coherence time. In ESR-STM experiments, Hahn echo sequences performed on single Ti atoms on MgO/Ag(001) reveal an exponentially decaying signal with a reported coherence time \(T_2 \approx 200~\mathrm{ns}\)\cite{yang_coherent_2019}. This is substantially longer than the corresponding Ramsey or Rabi coherence times (\(T_2^{\mathrm{Rabi}} \approx 40~\mathrm{ns}\)) and was interpreted as an effective suppression of slow magnetic-field fluctuations.

The measured Hahn echo sequences in Fig.~\ref{fig:fig2}b exhibit an exponential decay on the nanosecond timescale, with increasing signal amplitude and decreasing decay time as the tunneling current is increased. The extracted decay rates $T_2^{-1}$ scale linearly with $I$ (Fig.~\ref{fig:fig2}c), consistent with previous ESR-STM studies\cite{willke_coherent_2021}. This behavior has commonly been interpreted as tunneling-current--induced scattering:  The inelastic scattering of tunneling electrons, which arrive at a rate $\Gamma = I/e$, induces spin flips of the FePc spin (e.g. $I = 1~\mathrm{pA}$ corresponds to 1 electron tunneling every $160 ~\mathrm{ns}$ on average). It is noteworthy that decoherence does not require relaxation and for high-spin systems such as Fe adatoms on MgO, it was found that the tunneling decreases $T_2$ while $T_1$ is still much longer\cite{willke_probing_2018}. However, for spin-1/2 systems (Ti adatoms, FePc molecules) the extracted values of $T_2^{\mathrm{echo}}$ were found to be close to $T_1$\cite{yang_coherent_2019,willke_coherent_2021}, and it was therefore concluded that for Hahn echo sequences $T_2^{\mathrm{echo}}$ approaches the maximum value:

\begin{equation}
T_{\mathrm{2,echo}}^{\max} = 2 T_1
\end{equation}
\\
\noindent since any relaxation event also constitutes a loss of phase coherence. Thus, reducing the tunneling current should further enhance \(T_2\). From the slope of the current dependence, we extract a probability of a decoherence event per tunneling electron of $P_{T_2} = (\Delta T_2^{-1}/\Delta I) \cdot e \approx 20\%$, consistent with values reported previously for FePc on MgO/Ag(001)~\cite{willke_coherent_2021}. This probability for spin relaxation induced by tunneling electrons is very effective due to the high transition matrix element for spin S=1/2 when interacting with tunneling electrons\cite{ternes_spin_2015}.  While this interpretation is internally consistent, it implicitly assumes that the observed Hahn echo signal arises from genuine phase-coherent refocusing.

To further probe the nature of the observed decay, we extend the Hahn echo protocol by implementing Carr--Purcell (CP) sequences, which apply a train of equally spaced refocusing $\pi$-pulses (Fig.~\ref{fig:fig2}d). Figure~\ref{fig:fig2}e shows the extracted coherence times $T_2^{\mathrm{CP}}$ as a function of the number of $\pi$ pulses. In conventional settings, increasing the number of refocusing pulses further suppresses low-frequency noise and leads to an enhancement of the coherence time that ultimately saturates at a value bounded by $2T_1$\cite{degen_quantum_2017,de_lange_universal_2010}.
Contrary to this expectation, we observe a nearly linear increase of $T_2^{\mathrm{CP}}$ with pulse number, without any indication of saturation. We reach $T_2^{\mathrm{CP}} \approx 650~\mathrm{ns}$ for $N = 16$.  In Ref.~\cite{willke_coherent_2021}, $T_1 \approx 150~\mathrm{ns}$ for FePc using a relaxometry scheme consistent with reported lifetimes of other $S=1/2$ systems on the same surface (e.g., Ti adatoms: 190 ns\cite{yang_engineering_2017}).  This apparent violation of the expected bound $T_2 \leq 2 T_1$ raises questions about the interpretation of both the Hahn echo and CP-measurements, and motivated the control experiments and alternative interpretation discussed in the following sections.

\FloatBarrier
\section{Control Experiments and Relaxometry Interpretation}
\begin{figure*}[t]
	\centering
	\includegraphics[width=144mm]{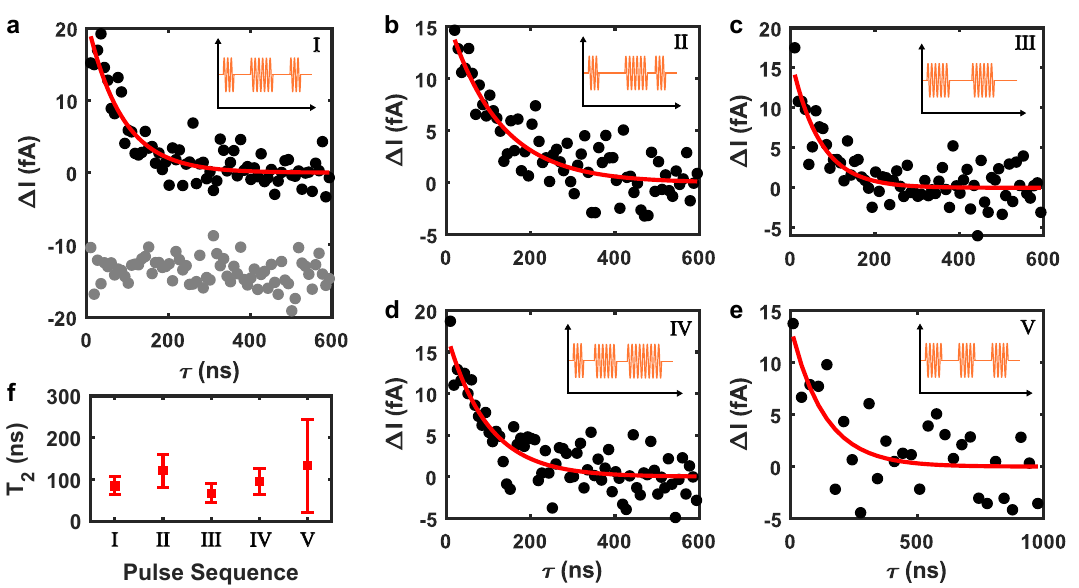}
  \caption{
  \justifying
  \textbf{Control experiments for Hahn echo measurements.} \textbf{(a)} Hahn echo sequence recorded on-resonance (black, $f=14.772~\mathrm{GHz}$) and off-resonance (gray, $f=15.4~\mathrm{GHz}$). A pronounced signal is observed only at the resonance frequency. The on-resonance trace yields an apparent echo decay time $T_2^{\mathrm{Echo}}=(85\pm22)~\mathrm{ns}$ from an exponential fit (red line), while the off-resonant dataset shows no signal ($I=5~\mathrm{pA}$, $V_{\mathrm{DC}}=40~\mathrm{mV}$, $V_{\mathrm{RF}}=60~\mathrm{mV}$, $B=536~\mathrm{mT}$,  $\tau_\pi=19~\mathrm{ns}$, $\tau_{\mathrm{cycle}}=2000~\mathrm{ns}$). \textbf{(b--e)} Control measurements testing the robustness of the exponential decay against imperfections in the pulse sequence. (b) unequal spacing between the two $\pi/2$ pulses and $\pi$ pulse; (c) use of two $\pi$ pulses instead of a $\pi/2$--$\pi$--$\pi/2$; (d) incorrect pulse length for the final $\pi/2$ pulse; (e) three pulses of equal length $2\pi/3$. Insets illustrate the respective pulse schemes.  [(b-d): $I=5~\mathrm{pA}$, $V_{\mathrm{DC}}=40~\mathrm{mV}$, $V_{\mathrm{RF}}=60~\mathrm{mV}$,  $B=536~\mathrm{mT}$, $f=14.772~\mathrm{GHz}$, $\tau_\pi=19~\mathrm{ns}$, $\tau_{\mathrm{cycle}}=2000~\mathrm{ns}$; (e): $I=5~\mathrm{pA}$, $V_{\mathrm{DC}}=60~\mathrm{mV}$, $V_{\mathrm{RF}}=60~\mathrm{mV}$, $B=523~\mathrm{mT}$, $f_0=13.813~\mathrm{GHz}$, $\tau_\pi=15.8~\mathrm{ns}$, $\tau_{\mathrm{cycle}}=2500~\mathrm{ns}$]. \textbf{(f)} Extracted $T_2^{\mathrm{Echo}}$ times from the measurements shown in (a)-(e). All datasets exhibit similar exponential decay behavior.}
	\label{fig:fig3}
\end{figure*}

As a first control experiment, we detuned the RF drive frequency $f$ away from the resonance frequency $f_0$  in the Hahn echo sequence (Fig.~\ref{fig:fig3}a). Off resonance, the signal vanishes within the noise floor, whereas on resonance a clear exponentially decaying response is recovered. This shows that the observed signal is linked to resonant spin transitions and is not caused by purely electrical rectification or instrumental artifacts.

We next tested whether the observed exponential decay is uniquely tied to phase-coherent refocusing, as expected for a genuine Hahn echo. To this end, we deliberately introduced imperfections into the pulse sequence (Fig.~\ref{fig:fig3}b--e), including unequal pulse spacings (Fig.~\ref{fig:fig3}b), incorrect durations for the final pulse (Fig.~\ref{fig:fig3}d), and pulse trains that do not implement a $\pi/2$--$\pi$--$\pi/2$ echo condition (Fig.~\ref{fig:fig3}c,e). Strikingly, all of these deliberately flawed sequences still yield a similar exponentially decaying signal (Fig.~\ref{fig:fig3}f). The presence of the decay under conditions that should suppress or eliminate coherent refocusing indicates that the dominant contribution to the observed signal does not rely on a true phase-coherent echo.

These observations motivate an alternative interpretation in which the measured decay reflects a relaxometry-like response. The key point is that in ESR-STM the applied RF voltage $V_{\mathrm{RF}}$ generates additional tunneling electrons that not only probe and relax the spin state: As discussed in the context of the Rabi oscillations in Eq.~\eqref{eq:deltaI}, the contribution proportional to $k_{\mathrm{RF}}$ enables spin-dependent detection via RF-induced tunneling currents -- even in the absence of a well-defined spin phase. Consequently, a pulse sequence nominally designed to refocus the coherent spin can -- under realistic ESR-STM conditions -- act primarily as a sequence that excites and probes the spin population.

Figure~\ref{fig:fig4} sketches the mechanism by which echo-like sequences can produce an apparent \textit{echo decay} that is governed predominantly by spin relaxation. Conceptually, the sequence can be viewed as an initial population preparation (first  $\pi/2$) followed by two population-sensitive probes ($\pi$ pulse and second $\pi/2$  pulse) separated by delay intervals:

\begin{figure}[t]
	\centering
	\includegraphics[width=88mm]{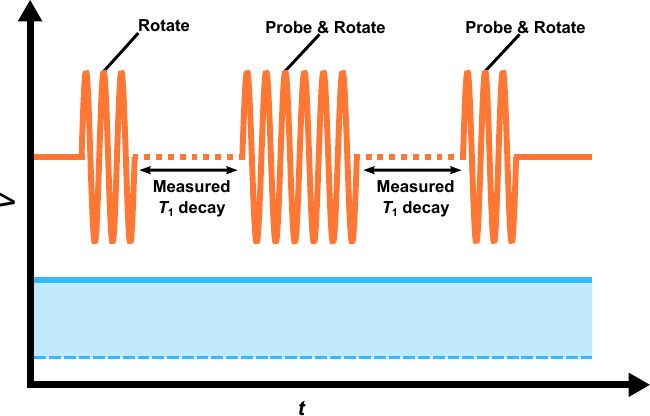}
  \caption{
  \justifying
  Schematic illustration of how Hahn echo pulse sequences can effectively act as two consecutive $T_1$ relaxation probes (orange: RF voltage; blue: DC voltage): The initial pulse rotates the spin system, while subsequent pulses -- intended for spin refocusing -- simultaneously probe and rotate the spin population. Each pulse thus partially measures the remaining magnetization, which produces an apparent exponential decay. This decay reflects $T_1$ relaxation rather than true spin decoherence ($T_2$). This leads to an echo-like decay signal whose exponential decay can be misinterpreted as a coherent process.}
	\label{fig:fig4}
\end{figure}

\begin{enumerate}
  \item[(i)] \textit{Population-sensitive detection during RF pulses.}
Each RF pulse generates tunneling currents that are spin-state dependent. Through the homodyne-like detection channel (captured by $k_{\mathrm{RF}}$), the junction response is sensitive to the present spin population. Importantly, this does not require phase coherence: A difference in current $\Delta I$ is already obtained if the spin state is different. Therefore, the later pulses (the $\pi$ pulse and second $\pi/2$  pulse of the Hahn echo) not only implement spin rotations, but also act as measurement pulses that partially read out the remaining longitudinal magnetization on a timescale set by $T_1$.

 \item[(ii)]\textit{Repeated excitation of the spin.}
Even if transverse coherence is lost between pulses, subsequent RF pulses still   drive transitions and generate spin-dependent tunneling currents. In   particular, once the spin has relaxed back toward the ground state $\lvert 0 \rangle$ during a delay interval, the following pulse again redistributes the population and probes the spin through the same RF-induced detection channel. As a result, the overall signal contains two consecutive $T_1$ relaxation responses within a total Hahn echo sequence.

  \item[(iii)] \textit{Relaxation driven by tunneling electrons.}
Still, the observed timescales and their dependence on the tunneling current are consistent with relaxation induced by tunneling electrons: This mechanism was discussed above as a source of $T_1$-limited decoherence and is supported by the current-dependent measurements (Fig.~\ref{fig:fig2}c and Fig.~\ref{fig:fig3}). However, even though the exponential decay extracted from the Hahn echo sequence is dominated by incoherent relaxation dynamics reflecting $T_1$, a genuine transverse coherence involving $T_2$ may decay on a shorter timescale.

\end{enumerate}

Thus, since successive pulses both drive and probe the spin population, the measured signal can acquire an exponential dependence during the delay times between pulses. This decay is determined primarily by  (longitudinal) relaxation rather than decoherence. In this regime, the extracted time constant resembles an echo decay but effectively reflects $T_1$-dominated dynamics.

This picture also provides a natural explanation for the behavior observed in the Carr--Purcell measurements presented in Fig. \ref{fig:fig2}d,e. Each of the $N$ refocusing pulses rotates and probes the spin population during a delay of duration $\tau/N$ between pulses and the coherence time is obtained via $T_2^{\mathrm{CP}}=N\cdot (\tau/N)$.  Thus, if each $\tau/N$ delay probes only the $T_1$ relaxation time and each pulse introduces again a spin rotation, the expression for $T_2^{\mathrm{CP}}$ is multiplied artificially to higher and higher numbers. As a result, the extracted coherence time can appear to grow with pulse number, without exhibiting the saturation expected for true dynamical decoupling. Each sequence segment still reflects a $T_1$-type relaxation, such that the apparent decay time increases with pulse number.
Taken together, the control experiments demonstrate that an exponentially decaying signal observed under a  Hahn echo protocol, such as that shown in Fig.~\ref{fig:fig2}b, is not by itself sufficient evidence for phase-coherent refocusing in ESR-STM.

\FloatBarrier
\section{Two-Delay Hahn Echo Protocol}
A defining characteristic of a genuine Hahn echo is the reliance on phase-coherent spin evolution during both free-precession intervals. While Hahn echo protocols presented so far vary both interpulse delays simultaneously, such measurements alone cannot unambiguously distinguish coherent refocusing from incoherent relaxometry-like responses in the given ESR-STM experiment. A more stringent test of coherence is provided by Hahn echo measurements with independently varied free-precession intervals\cite{childress_coherent_2006}.

In such a Hahn echo protocol, the first free-precession interval $\tau_1$ and the second interval $\tau_2$ are swept independently. For a phase-coherent spin system, an echo signal emerges only when the accumulated phases during the two intervals are properly refocused, resulting in interference features that depend on the relative values of $\tau_1$ and $\tau_2$. In contrast, a purely population-based or relaxometry-like signal does not require phase coherence and is therefore expected to depend only weakly on the relative timing of the two delays. 
Such delay sweeps have previously been demonstrated for atomic Ti spins in ESR-STM, but were limited to short delay times ($\tau_1 = \tau_2 \approx 20~\mathrm{ns}$)\cite{yang_coherent_2019}.  This only demonstrates that spin coherence is present after the given timescale $\tau_1 = \tau_2$ and that measurements should be performed on the scale of the presumed $T_2^{\mathrm{Hahn}}$ ($\approx 200~\mathrm{ns}$ for Ti adatoms). In the present work, two-delay echo measurements could not be reliably implemented for FePc, as no measurable signal was obtained.\\

\begin{figure*}[t]
	\centering
	\includegraphics[width=136mm]{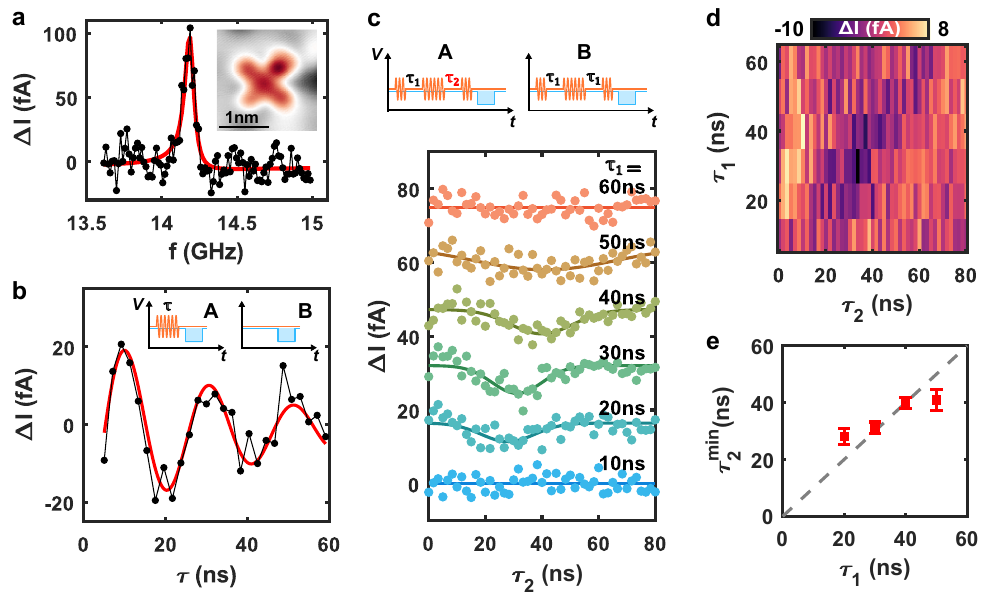}
  \caption{
  \justifying
  \textbf{Coherent spin control in an Fe--FePc S=1/2 molecular complex.} \textbf{(a)} ESR-STM spectrum of the Fe--FePc complex, showing a distinct resonance peak near $14.15~\mathrm{GHz}$. The inset displays an STM topography of the complex ($I=4~\mathrm{pA}$, $V_{\mathrm{DC}}= -60~\mathrm{mV}$, $V_{\mathrm{RF}}=30~\mathrm{mV}$, $B=461~\mathrm{mT}$). \textbf{(b)} Rabi oscillation measurements ($I=4~\mathrm{pA}$, $V_{\mathrm{DC}}= -60~\mathrm{mV}$, $V_{\mathrm{RF}}=60~\mathrm{mV}$, $B=461~\mathrm{mT}$, $f_0=14.15~\mathrm{GHz}$, $\tau_\pi=15.9~\mathrm{ns}$, $\tau_{\mathrm{cycle}}=300~\mathrm{ns}$, $\tau_{\mathrm{Probe}}=150~\mathrm{ns}$, $V_{\mathrm{Probe}}=-60~\mathrm{mV}$). The red curve represents a sinusoidal fit to the data yielding $\Omega=(48\pm4)~\mathrm{MHz}$. The inset illustrates the used pulse schemes in A and B cycle with a DC readout pulse instead of a constant DC background. \textbf{(c)} Two-dimensional Hahn echo measurements performed by varying the first delay time $\tau_1$ (10--60 ns) while sweeping the second delay $\tau_2$. The corresponding pulse schemes for lockin A and B cycles are shown at the top. A dip in signal is observed when $\tau_1$ and $\tau_2$ are equal, i.e. under echo conditions ($I=4~\mathrm{pA}$, $V_{\mathrm{DC}}= -60~\mathrm{mV}$, $V_{\mathrm{RF}}=60~\mathrm{mV}$, $B=461~\mathrm{mT}$, $f_0=14.26~\mathrm{GHz}$, $\tau_\pi=11~\mathrm{ns}$, $\tau_{\mathrm{cycle}}= 450~\mathrm{ns}$, $\tau_{\mathrm{Probe}}=150~\mathrm{ns}$,  $V_{\mathrm{Probe}}=-60~\mathrm{mV}$). The traces were vertically shifted for clarity. \textbf{(d)} Colormap of the full $\tau_1$--$\tau_2$ dataset, revealing the evolution and decay of the interference pattern characteristic for a coherent echo response. \textbf{(e)} Extracted echo position $\tau_2$ as a function of the first delay $\tau_1$, showing a near-linear relationship. Gray line is a guide to the eye with $\tau_1=\tau_2$}
	\label{fig:fig5}
\end{figure*}

To nevertheless establish an unambiguous benchmark for coherence, we performed two-delay Hahn echo measurements on Fe--FePc molecular complexes, which exhibit improved performance in coherent control protocols due to partial protection against tunneling electron-induced scattering\cite{huang_quantum_2025}. Figure~\ref{fig:fig5}a shows the ESR spectrum of a Fe--FePc complex, while Fig.~\ref{fig:fig5}b demonstrates coherent spin driving via Rabi oscillations.

In order to improve the sensitivity of the experiment, we optimize the pulse sequence shown in Fig.~\ref{fig:fig5}c (top). In the A cycle, the first delay time $\tau_1$ is kept fixed while the second delay $\tau_2$ is swept. In the B cycle both delay times are set equal to $\tau_1$. This ensures  identical total RF-pulse durations in both cycles and suppresses contributions from rectification. Moreover, since the relaxation processes discussed above are present in both cycles, the measurement becomes sensitive exclusively to the true echo signal. In addition, we here use a DC readout pulse at the end of the pulse sequence instead of a continuous readout voltage (see also \ref{fig:fig5}b). This ensures that the readout of the spin state is probed immediately following the coherent manipulation and prevents temporal overlap between RF and DC pulses, which further reduces current rectification.
Figures~\ref{fig:fig5}c and~\ref{fig:fig5}d present the results of the two-delay Hahn echo measurements, where $\tau_1$ is varied between $10~\mathrm{ns}$ and $60~\mathrm{ns}$ while $\tau_2$ is continuously swept. At short delay times, the data reveal dip features that depend on both $\tau_1$ and $\tau_2$. 

The evolution of the echo is further quantified in Fig.~\ref{fig:fig5}e, which shows the extracted echo position of  $\tau_1$ for different $\tau_2$. The approximately linear relationship between $\tau_2$ and $\tau_1$ is consistent with coherent phase accumulation during the first free-precession interval and subsequent refocusing by the pulse sequence. Notably, the coherent features decay rapidly and vanish for total evolution times exceeding approximately $40~\mathrm{ns}$. This indicates a limited intrinsic spin coherence time in this system. We attribute deviations from the perfect linear agreement $\tau_2=\tau_1$ (indicated in the plot) to the overall weak signal. 

Taken together, these measurements demonstrate that two-delay sweeps provide a good criterion for identifying genuine Hahn echo detection in ESR-STM. While one-delay echo decays can arise from incoherent relaxation processes mediated by tunneling electrons, the observation of echo interference patterns in $\tau_1$--$\tau_2$ space constitutes more robust evidence of phase-coherent spin dynamics.

\FloatBarrier
\section{Conclusion}
In this work, we have examined the interpretation of Hahn echo and Carr--Purcell pulse sequences in ESR-STM. While echo-based protocols are widely used to extract spin coherence times, our results demonstrate that, in ESR-STM, such measurements can be strongly influenced by tunneling electrons generated by the applied radio-frequency voltage. These electrons not only drive the spin system but also probe and relax it, giving rise to echo-like signals that do not necessarily rely on phase-coherent refocusing.

Through a series of control experiments, we showed that exponential decay signals persist even when the pulse sequence no longer fulfills the conditions for a coherent Hahn echo. This behavior is naturally explained by a relaxometry-like mechanism, in which successive RF pulses perturb and probe the spin population, leading to decay dynamics governed predominantly by the longitudinal relaxation time $T_1$. Within this framework, decay times extracted from conventional one-delay Hahn echo and CP measurements can significantly overestimate $T_2$.

We further demonstrated that two-delay Hahn echo measurements, in which both free-precession intervals are varied independently, provide a more stringent test of coherence. The observation of echo interference patterns in $\tau_1$--$\tau_2$ space constitutes clear evidence of phase-coherent spin evolution, whereas their absence indicates a dominant contribution from incoherent relaxation processes. Applying this criterion, we find that the intrinsic coherence times are comparable to those extracted from Rabi oscillations and substantially shorter than values inferred from one-delay echo protocols. Another route to establish more reliable pulse schemes is the use of a dispersive, remote readout\cite{wang_atomic-scale_2023}, in which the tunneling current primarily interacts with the readout spin.

More broadly, our results establish practical guidelines for the reliable interpretation of pulsed ESR-STM experiments. In tunneling-based spin detection schemes, the presence of an exponentially decaying echo signal alone is not sufficient to demonstrate coherent refocusing. Independent verification of phase coherence, for example through two-delay sweeps, is essential to distinguish genuine spin echoes from tunneling-induced relaxometry signals. These considerations are expected to be broadly relevant for future investigations of spin coherence in atomic and molecular systems using ESR-STM.

\newpage
\section*{Acknowledgements}
{P.W. acknowledges funding from the Emmy Noether Programme of the Deutsche Forschungsgemeinschaft (DFG, WI5486/1-2). P.G. and P.W. acknowledge financial support from the Hector Fellow Academy (Grant No. 700001123). P.W. and K.H.A.Y. acknowledge support from the Center for Integrated Quantum Science and Technology (IQST). P.W. acknowledges funding from the ERC Starting Grant ATOMQUANT. C.W., S.P, and A.J.H. acknowledge support from the Institute for Basic Science (IBS-R027-D1).}

\section*{Author contributions}
{P.G., W.H., and P.W. conceived the experiment. P.G., W.H., M.S., K.H.A.Y and P.W. set up the experiment and conducted the measurements. P.G., W.H., M.S., K.H.A.Y., C.W., S.P., A.J.H. and P.W. discussed and analyzed the experimental data. P.G. and P.W. wrote the manuscript with input from all authors. P.W. supervised the project.}

\section*{Data availability}
The data that support the findings of this study are available from the corresponding author upon reasonable request.

\printbibliography[title={References}]

\end{document}